IAC-10-D4.2.7

# ENERGY, INCESSANT OBSOLESCENCE, AND THE FIRST INTERSTELLAR MISSIONS

## M. G. Millis

Tau Zero Foundation, Fairview Park, United States, marc@tauzero.aero

Projections for the earliest interstellar mission possibilities are calculated based on 27 years of data on historic energy trends, societal priorities, required mission energy, and the implications of the Incessant Obsolescence Postulate (Where newer probes pass prior probes). Two sample missions are considered: launching a minimal colony ship where destination is irrelevant, and sending a minimal probe to Alpha Centauri with a 75 year mission duration. The colony ship is assumed to have a mass of $10^7$ kg, and the probe $10^4$ kg. It is found that the earliest interstellar missions could not begin for roughly another 2 centuries, or 1 century at best. Even when considering only the kinetic energy of the vehicles without any regard for propellant, the colony ship cannot launch until around the year 2200, and the probe cannot launch until around 2500. Examination of the Incessant Obsolesce Postulate shows that it becomes irrelevant under several conditions.

## I. INTRODUCTION

The situation that prompted this re-examination of humanity's interstellar readiness is having over a quarter century of data on human energy trends and space activity [1, 2]. From these sources, extrapolations are made for future interstellar missions.

Whereas past studies gauged humanity's readiness in terms of the doubling period of achievable speed [3, 4] or financial prowess [5], this study uses *energy* as its central variable. Energy is chosen because it is the fundamental currency of all physical transactions. Interstellar missions can be expressed as a function of the energy of motion. Higher energy capacities yield shorter trip times or greater carrying capacity. The energy required for various interstellar missions is compared to predictions of when those energy levels will be accessible to humanity.

Technology trades are not part of this study. With calculated energy readiness still in the distant future, it is premature to debate which propulsion and power options are best. The only exception is that multiple mission options are examined to reflect a span of possibilities.

Another factor considered when estimating the first launch is the *incessant obsolescence postulate* [6: p.157] and other perspectives. Incessant obsolescence is the notion that any interstellar mission will be passed by more advanced technology launched later, which discourages the launching of any interstellar mission. Since such expectations impede progress, they are examined.

## II. ENERGY RESOURCES

When estimating the amount of energy that humanity can devote to interstellar missions, two factors are considered: the total amount of energy produced by all humanity and the proportion of that energy which is devoted to space endeavors. Now that 27 years of data exist, reasonable energy production trends can be inferred for use in extrapolations, including deviations of those trends to reflect the degree of uncertainty. Also, using 27 years of data on the Space Shuttle missions, comparisons can be inferred regarding the proportion of energy that society devotes to major space missions.

### Trends of Human Energy Prowess

Annual data of the world's total energy production, over the period from 1980 through 2007, is used to calculate energy growth trends [1]. This includes energy produced via petroleum, natural gas, coal, and electric power. Electric power includes hydroelectric dams, nuclear generators, geothermal sources, solar and wind power, and other minor sources. This data can be presented in terms of energy *per year* (Joules), or in terms of power (Watts), where that conversion includes $3.2 \times 10^7$ sec/yr.

As an aside, it is interesting to note that the units used in the "U.S. Energy" data are in the units of *Quadrillion Btu's*, where the United States definition of Quadrillion is $10^{15}$, yet the United Kingdom definition is $10^{25}$. This can be quite confusing since these data are from the United States, reported in terms of *British Thermal Units*. This peculiarity is raised here to prevent other researchers from incorrectly converting the data. The correct conversion is: Quadrillion = $10^{15}$.

The ratio of each year's energy production to its preceding year is calculated (starting with 1981 growth), then the average and standard deviation of all these 27 years are calculated. It is found that the average growth rate in world energy production has been 1.89% ± 1.73% (error band is *one* standard deviation).

Note the wide span of uncertainty in this result. With such uncertainty, there is no advantage in refining the remaining calculations to a high level of precision. To help convey this uncertainty, the final answers will be bracketed by high and low variations of one standard deviation of the average growth rate.

As much as one can assert that such energy projections might be invalidated by depletion of natural resources, one could equally assert that unforeseen advances will discover new energy sources, leaving the growth trends relatively stable. At least for these order-of-magnitude estimates, a constant growth rate is therefore assumed, and uncertainty estimates are based on the standard deviations of the historical annual growth rates.

To set this in context, **Table-1** shows how this predicted span of growth compares to noteworthy thresholds of human energy prowess, such as the Kardashev Scale [7]. The calendar years are calculated by comparing the 2007 total energy value, growth rates, and the specific energy thresholds using Eq. 1:

$$\text{Year} = 2007 + \log_{(\text{rate}+1)} \frac{(\text{Desired power level})}{(\text{Power level in 2007})} \quad [1]$$

Where:
Year = Calendar year when the desired power level is achieved
rate = Annual power production growth rate

| Power Accessible by Humanity | Calendar Year Predictions | | |
|---|---|---|---|
| | Earlier | Nominal | Later |
| Starting Baseline $1.6 \times 10^{13}$ W $= 5.0 \times 10^{20}$ Joules/yr | – | **2007** | – |
| MOON Solar flux $1.6 \times 10^{15}$ W | 2137 | **2253** | 4888 |
| Kardashev, Type I* EARTH Solar flux $2.1 \times 10^{16}$ W | 2209 | **2390** | 6498 |
| Kardashev, Type II SUN total Solar flux $3.8 \times 10^{26}$ W | 2873 | **3652** | 21272 |
| Kardashev, Type III GALAXY power $4 \times 10^{37}$ W | 3587 | **5007** | 37147 |
| Based on the following span of annual growth rates (1980-2007) | 3.62% Ave + standard deviation | **1.89% Average** | 0.16% Ave – standard deviation |

**Table 1: Extrapolating Humanity Energy Prowess**

*Regarding the Kardashev scales, this assessment takes exception to the 1964 value specified by Kardashev for *Type I* civilizations. In principle his definition states that the civilization has mastered the total power of its planet. Kardashev judged Earth to have achieved that level in 1964 – with a total power of $4 \times 10^{12}$ W. Instead, this assessment accepts Kardashev's *definition*, but defines the requisite energy as identical to all the energy reaching Earth from the Sun. Based on an average solar flux of 164W/m² on projected surface area of $1.3 \times 10^{14}$ m² (flat disk of Earth radius perpendicular to Sun), this equates to $2.1 \times 10^{16}$ W. Per this paper, humanity will not achieve a *Type I Kardashev* status until its annul power production is $2.1 \times 10^{16}$ W. As shown in **Table-1**, this is not achieved until the year 2200 at the earliest, or nominally, not until 2400 (rounding to nearest century).

To provide a nearer-term significant milestone for comparison, the total solar flux on the Moon is offered as an earlier target, and is calculated to be $1.6 \times 10^{15}$ W (based on flux of 164W/m² on a projected surface area of $9.5 \times 10^{12}$ m²). Astronomical values used in these calculations are from reference 8.

Portion Devoted to Spaceflight

The amount of resources devoted to spaceflight is more of a sociological phenomenon than technological. For a realistic estimate, annual Space Shuttle launch rate [2] is compared to the total annual energy consumed by the United States [1] over the years 1981 to 2007.

The comparison to the *United States* energy instead of *world* energy is because the Shuttle was only a United States investment. When extrapolating to future investments in interstellar missions, however, the assumption is that this same proportion will be applied on a world-wide scale, since *interstellar* missions impact all of humanity.

Although the Space Shuttle does not represent the *total* devotion to space activities in the United States, it is considered a comparable, major space activity. In much the same way that other missions occurred during the Shuttle era, it is reasonable to suspect that such multi-tasking will also exist in the future. Accordingly it is sufficient to consider Shuttle data alone to derive this devotion ratio.

Another detail is that only the *propulsive* energy of the Shuttle is used. This is because subsequent comparisons will also be in terms of the propulsive energy of interstellar flight. The Shuttle launch energy of $1.2 \times 10^{13}$ Joules-per-launch is from a calculation in *Frontiers of Propulsion Science* [9: p.147].

Based on these assumptions and data, it is found that the maximum ratio of Shuttle propulsion energy to total US energy consumed occurred in the year 1985, and equals $1.3 \times 10^{-6}$. The average ratio over the years 1981 to 2007, is $5.5 \times 10^{-7}$.

Note that the difference between the *average* and *maximum* ratio is over an order of magnitude. For predicting the *earliest* opportunities of future missions, the *maximum* proportion is. This ratio, dubbed, *Space Devotion Ratio*, $R_{SD}$ is taken to be **$1.3 \times 10^{-6}$**.

## III. INTERSTELLAR ENERGY ESTIMATES

To provide a reasonable mix of possibilities, two different scenarios are examined for "first launch" opportunities: a minimum colony ship and "lifespan duration" probe (75-yr). Both of these are anticipated to be lower energy examples that could be achieved sooner than more ambitious missions. Two propulsion energies are calculated for each scenario, an ideal case and an advanced rocket option. Then the corresponding world energy levels to enable those missions are calculated.

Propulsion Options

The propulsion options that are analyzed span the most ideal case (where only the kinetic energy of the vehicle is considered) and some form of advanced rocket.

The kinetic energy version provides the lowest possible estimate by assuming 100% efficient conversion of stored energy into kinetic energy of the vehicle. Whether this is enabled by some form of perfect beamed-energy propulsion or even the discovery of a breakthrough space drive, is irrelevant. What matters is that this value cannot be bested by hoping for technological improvements in propulsion.

To provide a representative estimate in the other direction, but still aiming for promising performance, a rocket having an exhaust velocity of 0.03c is assumed (3% light-speed, which is equal to $9.0 \times 10^6$ m/s, and an $I_{sp}$ of $9.2 \times 10^5$ s). This value is consistent with various nuclear propulsion notions recently compiled by Robert Frisbee [9: pp. 77, 78].

It should be noted that this is not the highest performance covered in Frisbee's assessments. The other option is an order of magnitude higher, crossing into the relativistic regime of 0.3c (30% of light-speed, which is equal to $9.0 \times 10^7$ m/s, and an $I_{sp}$ of $9.2 \times 10^6$ s). This was based on extremely optimistic projections of antimatter-matter annihilation [9: p. 73]. For the assessment in this paper, where it is desired to span a range of ideal to modest performance, this highest value seems inappropriate.

Minimum Colony Ship

The simplest scenario to calculate is the launching of a colony ship to ensure humanity's survival from threats to Earth's habitability. Since there is no destination, only the escape velocity and ship mass need to be estimated to determine the propulsion energy. For the sake of estimating *earliest* launch possibilities, the following *minimum* values will be used in estimating its requirements:
- Population = 500
- Mass per person = 50 Metric Tons ($5 \times 10^4$ kg)
- Resulting *total* ship dry mass, $m = 2.5 \times 10^7$ kg
- Onboard power per person = $1.1 \times 10^4$ W
- Resulting *total* ship power = $5.6 \times 10^6$ W
- Required $\Delta v = 4.2 \times 10^4$ m/s
- Propulsion characteristics assessed:
  - Propellantless (Kinetic energy equation)
  - Advanced rocket ($v_{ex} = 9.0 \times 10^6$ m/s)

The estimate for the required population starts with John Moore's assessment that the absolute minimum number of individuals for a self-perpetuating colony is 180 [10]. That minimum is increased by a modest factor for margin, but is far less than O'Neill's space colony, which was envisioned to hold 10,000 people [11].

The mass factor chosen is a low rate of 50 Metric Tons per person, the same assumption used by O'Neill [11]. For comparison, Salkeld suggests 100 MT [12], and Cassenti uses 500 MT [4].

The power-per-person estimate is from the actual power-per-person consumed in the United States in 2007 (latest data). The onboard power estimate is provided as a check to see if it is comparable to the propulsion power, to determine if it must be included in the estimate calculations. Since it turns out a few orders of magnitude less than the propulsion energy, it will have no bearing on the total energy calculations. The power level is still listed, however, for reference.

The required delta-v ($\Delta v$) is based on the "escape velocity" from the solar system and is calculated by setting kinetic energy of the vehicle equal to the gravitational potential energy at the radius of departure (Earth's solar orbit), and then solving for velocity:

$$v_{\text{escape}} \geq \sqrt{\frac{2MG}{R}} \quad [2]$$

Where these values are used [8]:
G = Gravitational constant = $6.7 \times 10^{-11}$ m$^2$/kg•s$^2$
M = Mass of Sun = $2.0 \times 10^{30}$ kg
R = Radius of Earth's solar orbit = $1.5 \times 10^{11}$ m

Note that the answer of $4.2 \times 10^4$ m/s is non-relativistic, therefore non-relativistic equations can be used to assess this minimal colony ship scenario.

While it can be argued that such a mission would not actually need to leave the solar system, the chosen delta-*v* is considered a reasonable initial estimate. Regardless of final trajectory, some degree of delta-*v* capability would be required.

The resulting energy for the ideal propellantless propulsion option is calculated simply by kinetic energy (Eq. 3) and found to be **$2.3 \times 10^{16}$ Joules**.

$$KE = \frac{1}{2}mv^2 \quad [3]$$

Where:
m = Mass of ship = $2.5 \times 10^7$ kg
v = Escape velocity = $4.2 \times 10^4$ m/s

The propulsion energy for the rocket option is calculated using Eq. 4, which is derived from the burn-out velocity form of the rocket equation, which is then converted into terms of energy [9: p. 145]. This results in a required launch energy of **4.8 x10^18 Joules**.

$$E = \frac{1}{2} m \left( e^{\left(\frac{\Delta v}{v_{ex}}\right)} - 1 \right) (v_{ex})^2 \quad [4]$$

Where:
- $m$ = Mass of ship = 2.5 x10^7 kg
- $\Delta v$ = Escape velocity = 4.2 x10^4 m/s
- $v_{ex}$ = Exhaust velocity = 9.0 x10^6 m/s

Seventy-Five-Year Interstellar Probe

The next scenario is sending a probe to our nearest-neighbouring star system, Alpha Centauri, within the most tolerable mission duration.

A duration of 75 years is chosen as the most extreme duration that might be tolerated by mission planners. Although prior studies suggested durations of 40 years [4] or 50 years [13], this study increases that patience to a 75-year mission duration when considering increases in human longevity and the development of tools for allowing continuity of mission information (e.g. Internet). This value is selected as a compromise between a long career span of 50 years, and an extended longevity of 100+ years. Also, 75 years still seems short enough to be within the limit of autonomous equipment reliability.

Another critical assumption is that the interstellar mission is a *rendezvous*, not a *flyby*. This is based on allowing enough time at the destination to acquire data worthy of the 75-year wait. For a distance of 4.3 ly and a 71 yr transit time (4 years of the 75 yr mission is for data to return to Earth), the speed past the destination would be 6% light-speed. To put this into perspective, this means that a fly-by probe would only be within ± 1 AU for less than 5 hours:

$$\text{Time on Target} = \frac{2 AU}{6\% c} = \frac{2(1.5 \times 10^{11} m)}{(0.06)\left(3 \times 10^8 \frac{m}{s}\right)\left(\frac{3600 s}{hr}\right)} = 4.6 \, hr \quad [5]$$

The probe mass is assumed to be a modest extrapolation of prior deep space probes, with the assumption that the additional mass would be required for design margins to endure the 71 year transit, plus allowing more substantial scanning and communication equipment to relay meaningful data back to Earth. For reference, **Table-2** presents comparisons of the mass and onboard power of historic probes that were sent to the outer regions of our solar system, with some continuing beyond.

| Probe Examples | Launch Year | Mass (kg) | Power (kW) |
|---|---|---|---|
| Voyager 1 & 2 | 1977 | 723 | 0.4 |
| Galileo | 1989 | 2,380 | 0.6 |
| Cassini-Huygens | 1997 | 5,600 | 0.8 |
| Deep Space 1 | 1998 | 374 | 2.5 |
| New Horizons | 2006 | 478 | 0.2 |
| Dawn | 2007 | 1,250 | 10 |

Table 2: Comparing Historic Probes

Another consideration is the power required on the probe to relay its findings back to Earth. Based on recent assessments, the onboard power requirement spans 100 W – 1000 kW, depending on whether optical or radio communication is used [9: p. 100]. Since these energy levels are not comparable to the propulsion demands, they have no bearing on the total energy calculations. They are listed, however, to answer other interests.

The resulting list of probe mission characteristics includes:
- Total dry mass, $m = 10^4$ kg
- Onboard power for communication = $10^2 – 10^6$ W
- Trip distance, $d$ = 4.3 ly = 4.1 x10^16 m
- Mission duration, $t_m$ = 75 years (incl. signal time)
- Transit time, $t_t$ = 70.7 years
- Average resulting transit velocity, $v_{ave}$ = 0.06c
- Required *increment* $\Delta v$ = 1.8 x10^7 m/s
- Propulsion characteristics assessed:
  - Propellantless (Kinetic energy equation)
  - Advanced rocket ($v_{ex}$ = 9.0 x10^6 m/s)

To proceed with the energy calculations it is assumed that the thrusting time is short relative to the entire mission duration. This allows the use of Eq's. 3 and 4 again. For rockets, this is a very reasonable assumption considering how quickly they consume their propellant.

Unlike the colony ship, this mission requires *two* impulses, the initial acceleration plus the deceleration when reaching the destination.

The energy for the propellantless version now requires that Eq. 3 is applied twice, once for acceleration and once for deceleration, and both of these use the *average* transit velocity to calculate the changes in kinetic energy. Following these methods and inputting the vehicle mass, the propulsion energy of the propellantless version is found to be **3.3 x10^18 Joules**.

The propulsion energy for the rocket version is again calculated with Eq. 4, but now where the delta-v is *twice* the average transit velocity; one increment of delta-v for acceleration, and one for deceleration. Following these methods, and inputting the vehicle mass, the energy of the rocket version is found to be **2.3 x10^19 Joules**.

Converting Propulsion Energy to Human Readiness

To determine when humanity will be able to launch an interstellar mission, the missions' propulsion energy is converted into equivalent world energy values using Eq. 6, which uses the Space Devotion Ratio calculated previously:

$$P_{World} = \frac{E_{Prop}}{t \cdot R_{SD}} \quad [6]$$

Where:
- $P_{World}$ = Equivalent world power level, Watts
- $E_{Prop}$ = Propulsion Energy, Joules
- $t$ = Time to acquire energy ≡ 1 yr = $3.2 \times 10^7$ s
- $R_{SD}$ = Space Devotion Ratio = $1.3 \times 10^{-6}$

The time to accumulate the propulsion energy is assumed to be one year to keep commonality with the *annual* data format of world energy predictions.

Alternatively, one could entertain the option of spreading that energy accumulation over a longer period to reduce the overall energy readiness date. Given the broad uncertainties already noted, the assessment option chosen here is to keep the analysis simple – using *annual* comparisons. It can be argued that spreading the mission energy demands across multiple years would only affect the readiness estimates by a few years, but the uncertainty of the final estimates spans centuries, or decades at best.

With those propulsion energies now converted to their equivalent world power levels, the calendar years when those levels might be first achieved are calculated using Eq. 1. The findings are presented in **Table-3**.

| World Power Level and Corresponding Interstellar Mission | Calendar Year Predictions | | |
|---|---|---|---|
| | Earlier | **Nominal** | Later |
| **Colony Ship,** $m = 2.5 \times 10^7$ kg, $\Delta v = 4.2 \times 10^4$ m/s | | | |
| Kinetic energy alone $5.5 \times 10^{14}$ W | 2106 | **2196** | 4220 |
| Advanced technology $1.2 \times 10^{17}$ W | 2257 | **2482** | 7571 |
| **75-yr Probe,** $m = 10^4$ kg, $\Delta v = \Delta v = 6.6 \times 10^7$ m/s | | | |
| Kinetic energy alone $8.1 \times 10^{16}$ W | 2247 | **2463** | 7343 |
| Advanced technology $5.6 \times 10^{17}$ W | 2301 | **2566** | 8551 |

**Table 3: Predicting Interstellar Mission Readiness**

From these it is evident that the sequence of earliest achievable missions – in terms of energy availability – is as follows:
- 1st  Non-propellant colony ship in 2200.
- 2nd  Non-propellant 75-yr probe in 2500, and (tie) Advanced rocket colony ship in 2500
- 3rd  Advanced rocket 75-yr probe in 2600.

Note the substantial span of uncertainty in these results in **Table-3**, which stems from the huge span in possible energy production growth rates. This is why the calendar dates, in the list above, have been rounded to the nearest century.

## IV. REFLECTING ON FINDINGS

It is interesting to compare the main findings of this study to prior studies. Again, this study found that the first interstellar mission does not appear possible for another 2 centuries. Using projections of vehicle speed improvements, Cassenti's 1982 estimate also concludes that 2 centuries remain [4]. And finally, using economical projections, Dyson's 1968 estimate also concludes that 2 centuries remain [5].

Although it may seem strange that the colony ship appears be achievable sooner than the small probe, recall that energy is proportional to the square of *velocity* and linearly proportional to *mass*. Thus, the probe, having a flight velocity 3-orders-of-magnitude greater than the colony ship concedes its advantage of being 3-orders-of-magnitude less massive than the colony ship.

Another factor worthy of deeper examination is the energy budget for the colony ship. At a certain point this energy will be substantial enough that it must be included in its readiness estimate. The longer that the colony ship is planned to survive on its own without recharging its energy reserves, then the more energy must be loaded aboard the ship at the time of launch. When this onboard energy becomes comparable to the propulsion energy, then it must be taken into account when estimating the launch readiness date. As shown in **Table-4**, a mere 100 year duration is comparable to the lowest launch energy (non-propellant case), whereas, that energy does not approach the rocket case until a 10,000 year operation is considered:

| Colony Ship Operational Duration (years) | Required On-Board Energy Reserves (Joules) |
|---|---|
| 100 | $2 \times 10^{16}$ |
| 1000 | $2 \times 10^{17}$ |
| 10000 | $2 \times 10^{18}$ |
| **Contrast to Propulsion Energy** | |
| Non-propellant propulsion | $2 \times 10^{16}$ |
| Rocket propulsion | $5 \times 10^{18}$ |

**Table 4: Comparing Colony Ship On-Board Energy to Propulsion Energy**

To view the findings together, **Fig. 1** plots comparisons based in the data compiled in **Table-5**. Note that the values only show 1 significant digit and the calendar years are rounded to the nearest century.

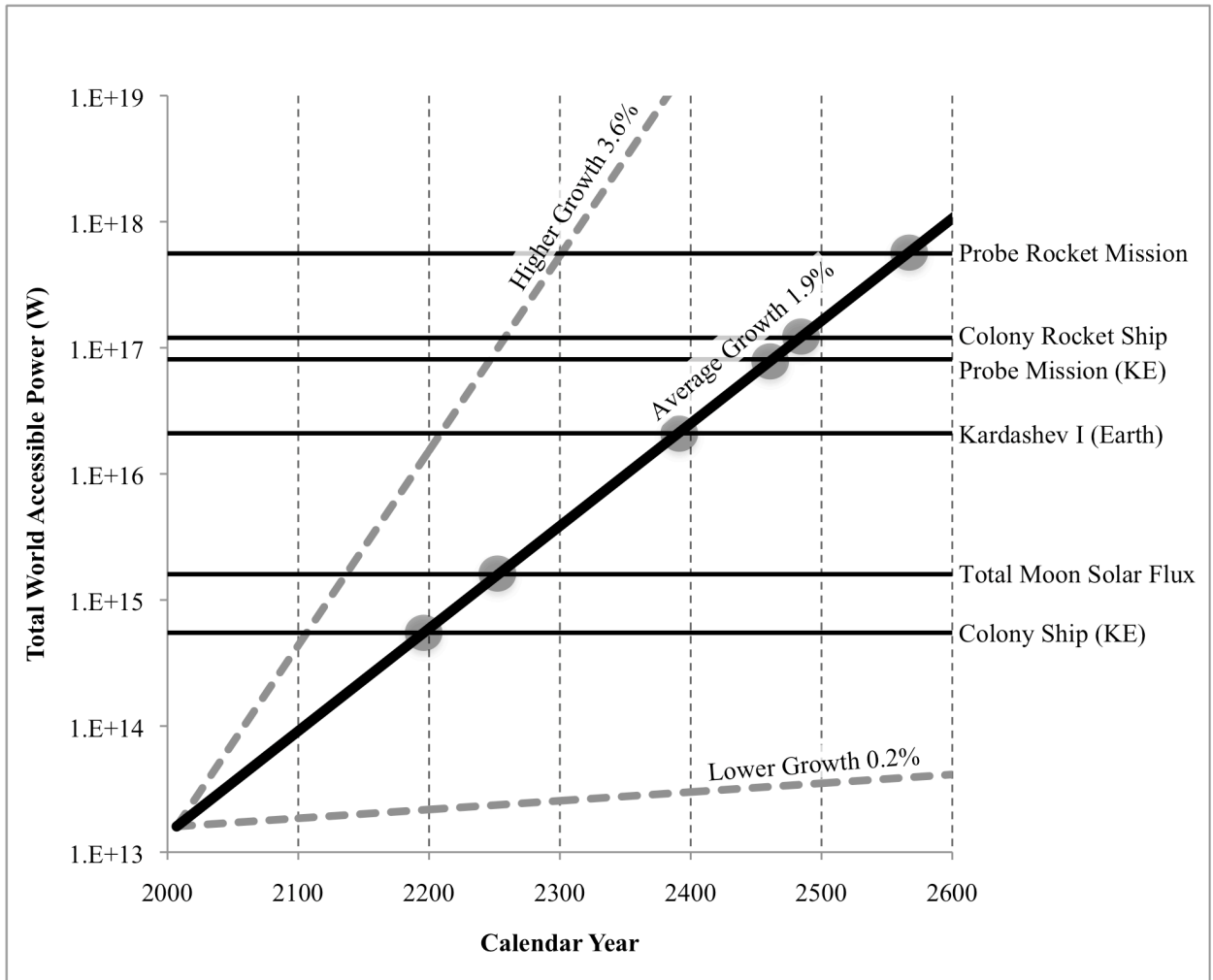

**Figure 1: Comparison of Humanity's Growing Power versus Required Interstellar Energy Milestones**
The horizontal energy thresholds shown for the Missions are in terms of the equivalent world power levels required to supply the requisite propulsion energy when considering that only $10^{-6}$ of the worlds energy is available to such missions.

| Readiness Date | | | Milestone | Total World Accessible Power (W) |
|---|---|---|---|---|
| Earlier 3.6% | Nominal 1.9% | Later 0.2% | | |
| – | **2007** | – | Starting Baseline ($5 \times 10^{20}$ Joules per year produced) | $2 \times 10^{13}$ |
| 2100 | **2200** | 4200 | Colony Ship Mission (*non-propellant*, E = $2 \times 10^{16}$ Joules) | $6 \times 10^{14}$ |
| 2100 | **2300** | 4900 | Capture 100% Solar Flux on Moon | $2 \times 10^{15}$ |
| 2200 | **2400** | 6500 | Kardashev Level I Achieved = 100% Solar Flux on Earth | $2 \times 10^{16}$ |
| 2200 | **2500** | 7300 | Probe Mission (*non-propellant*, E = $3 \times 10^{18}$ Joules) | $2 \times 10^{16}$ |
| 2300 | **2500** | 7600 | Colony Ship Rocket Mission ($v_{ex}$ = *3% c*, E= $5 \times 10^{18}$ Joules) | $1 \times 10^{17}$ |
| 2300 | **2600** | 8600 | Probe Rocket Mission ($v_{ex}$ = *3% c*, E= $2 \times 10^{19}$ Joules) | $6 \times 10^{17}$ |
| 2900 | **3700** | 21000 | Kardashev Level II Achieved = 100% Solar Flux Captured | $4 \times 10^{26}$ |
| 3600 | **5000** | 37000 | Kardashev Level III Achieved = 100% Galaxy Energy Captured | $4 \times 10^{37}$ |

**Table 5: Compilation of Findings – Predicting Readiness Years From Energy Comparisons**

## V. OTHER MISSION PROCRASTINATIONS

In addition to the energy availability limitations, other factors affect when interstellar missions are likely to be undertaken, including the incessant obsolescence postulate, motivations, and perceptions of practicality.

### Incessant Obsolescence Postulate

As much as continuing advances in science and technology will make it easier to launch an interstellar mission, these advances also create a quandary, the ***incessant obsolescence postulate*: No matter when an interstellar probe is launched, a subsequent probe will reach the destination sooner and with more modern equipment.** This is only a postulate, not a theorem nor even a principle. It is presented here not as an immutable constraint, but as one of the impediments for launching interstellar missions.

As an aside, the term, *incessant obsolescence postulate*, was first coined by the author around 1999 [6: p. 157]. This same notion has been called, the *incentive trap*, by Andrew Kennedy [3] and *Zeno's paradox in reverse*, by David Brin (a term possibly originating during the 1994 workshop; "Interstellar Robotic Probes – Are we ready?" [14]).

Although this incessant obsolescence postulate appears valid, it will eventually expire, as has been shown by Kennedy [3] and others (personal communication, Gerald Nordley, 2006-June). Due to the combination of the nonlinear nature of both advancement trends and relativistic spaceflight, there will be a point where an optimum launch opportunity occurs. Waiting longer does not get you to the destination sooner. Kennedy dubbed his optimization calculation the "*wait equation*" [3].

A version of that analysis in terms of energy, rather than technology development, was not completed in time for this paper. This remains a curiosity for future assessments. A suitable mission scenario for such an analysis would be sending a minimal crew and set of equipment to a begin colonizing a habitable planet, where minimal transit time is a critical factor.

In addition to the eventual expiration of incessant obsolescence, there are other conditions that would make this postulate collapse, such as:
- Significantly closer destinations become available (reduces trip time)
- Trip time becomes irrelevant:
  - Colony ships without specific destinations
  - Human lifespan increases dramatically
- The pace of technological development or energy production dramatically slows (societal retardation)
- Propulsion physics breakthroughs are achieved (significantly reducing trip times)
- Motivations other than *first-to-destination* drive mission planning.

### Implicit Motivations

The most significant factor that might negate the incessant obsolescence postulate is simply motivation. The incessant obsolescence postulate is only a limitation when the motivation is to *reach the destination first* – as if interstellar flight were a race. Instead, for example, if the motivation is to *depart* the Earth the earliest, then the incessant obsolescence postulate is irrelevant. If the goal is to stimulate technological developments, then launching soon and frequently is more prudent. If the purpose is to provide humanity with an alternate survival strategy, then the focus would be toward building a colony ship, where trip time is irrelevant.

To reflect the impact of differences in motivation, **Table-6** is offered:

| Motivations | Corresponding Focus |
|---|---|
| Being first: conquest of space | Stymied by the Incessant Obsolescence Postulate |
| Human survival beyond the habitability of Earth | Self-sustaining, multi-generation colony ships |
| Human expansion: finding and settling other habitable worlds | 1. Find where nearest habitable worlds exist. 2. Develop technology to send a colony there. |
| Incremental gains | Seeking returns greater than investment within tolerable duration |
| Intellectual curiosity: learning more about our place in the universe and expanding our abilities | Limited to discretionary resources and de-focused by multiplicity of interests |

**Table 6: Comparing Motivations & Actions**

### Perceptions Affecting Practicality

In addition to the influence of motivations, there is also an effect from perceptions of *practicality* – perceptions which can be implicit and thus unknowingly affect decisions. *Practical* is a relative term depending on one's priorities. To some this means choosing the most feasible approach, regardless of how long it takes to reach another star system. To others, star ships are not practical unless they are generation ships, and to still others, star flight won't be practical until it is as simple as envisioned in science fiction.

To introduce this span of perspectives **Table-7** compares facets of practicality to different levels of technological ambition. Consider, for example, if economy and feasibility were the driving factors. Although this would make it practical to launch something soon, only a tiny probe could be sent and it would take dozens of millennia just to reach our nearest neighboring star – certainly longer than the spacecraft's warranty. Conversely, when considering star-flight within a human lifetime, seeking breakthroughs in physics would seem practical.

|  | Feasibility | Time | Payload | Cost |
|---|---|---|---|---|
| **Available technology** | 100% | 10's of millennia | $10^3$ kg | $10's of millions |
| **Advanced technology** $v_{ex} = 0.03c$ | 80% | Centuries | $10^4$ kg - $10^6$ kg | $10's of billions |
| **Ultimate technology** $v_{ex} = 0.3c$ | 20% | Decades | $10^7$ kg | –?– |
| **New physics propulsion** | –?– | Months? (FTL) | $10^7$ kg + | –?– |

**Table 7: Differing Perspectives of Practicality**

## VI. CONCLUDING REMARKS

The recurring conclusion from this and two other cited studies is that interstellar missions still seem 2-centuries away. This is true despite the fact that each of these studies uses different methods of assessment: energy, technology, and finance.

Considering that recurring finding, it seems premature to attempt to focus on "best" propulsion options or to be inhibited by the spectre of the incessant obsolescent postulate.

With no single technical solution around the corner, it seems more prudent to hedge our bets across the whole span of ambitions, but with cycles of short-term, affordable investigations that target the critical questions whose answers can be sought today. This can span the seemingly simple solar sails all the way to the seemingly impossible faster-than-light travel.

There is a cliché that the journey is more important than the destination. Applying the analogy that the *destination* is to achieve the first interstellar mission, perhaps we should focus our attention on finding value in the *journey* – the value in the research steps that chip away at these unknowns and add to knowledge today.

Ad astra, incrementis – to the stars in steps, where each is greater than before.